\documentclass[twocolumn,aps,prd,showpacs,preprintnumbers]{revtex4}
\usepackage{graphics}
\usepackage{epsfig} 
\def\beq{\begin{equation}}
\def\eeq{\end{equation}}
\def\barr{\begin{eqnarray}}
\def\earr{\end{eqnarray}}

\def\lsim{\raise0.3ex\hbox{$\;<$\kern-0.75em\raise-1.1ex\hbox{$\sim\;$}}}
\def\gsim{\raise0.3ex\hbox{$\;>$\kern-0.75em\raise-1.1ex\hbox{$\sim\;$}}}

\def\dmsq{\Delta m^2}
\def\msqa{\Delta m^2_{\rm atm}}
\def\msqs{\Delta m^2_{\odot}}

\def\nue{{\nu_e}}
\def\nuebar{{\bar{\nu}_e}}

\def\la{\langle}
\def\ra{\rangle}

\begin{document}

\title{Phase effects in neutrino conversions during
a supernova shock wave}

\author{Basudeb Dasgupta and Amol Dighe}
\affiliation{Tata Institute of Fundamental Research,
Homi Bhabha Road, Mumbai 400005, India}

\begin{abstract}

Neutrinos escaping from a core collapse supernova a few seconds
after bounce pass through the shock wave, where they may
encounter one or more resonances corresponding to 
$\Delta m^2_{\rm atm}$.
The neutrino mass eigenstates in matter may stay coherent 
between these multiple resonances, giving rise to oscillations 
in the survival probabilities of neutrino species.
We provide an analytical approximation to these inevitable phase effects, that relates the density profile of the shock wave to the oscillation pattern.
The phase effects are present only if the multiple resonances
encountered by neutrinos are semi-adiabatic,
which typically happens for 
$10^{-5} \lsim \sin^2 \theta_{13} \lsim 10^{-3}$.
The observability of these oscillations is severely limited by
the inability of the detectors to reconstruct the 
neutrino energy faithfully. 
For typical shock wave profiles, the detection of
these phase effects seems rather unlikely.
However, if the effects are indeed identified in the $\nuebar$ spectra,
they would establish inverted hierarchy and a nonzero 
value of $\theta_{13}$.

\end{abstract}




\maketitle

\section{Introduction}                                   
\label{intro}

The neutrino fluxes emitted from a core collapse supernova (SN) 
contain information about the primary fluxes produced inside 
the star, the neutrino mixing pattern as well as the matter 
densities that the neutrinos have passed through.
The high statistics neutrino signal that one expects from
a future galactic SN needs to be decoded in order to extract this information.

The neutrinos and antineutrinos produced inside the SN pass through 
the core, mantle and envelope of the progenitor star before
escaping.
They encounter matter densities ranging from nuclear densities
deep inside the core all the way to vanishingly small densities
in the interstellar space.
Neutrino masses and mixing angles, and hence the extent of their
flavor conversions, depend crucially on the density of 
surrounding matter, hence it is important to study these matter
effects in detail.

The matter effects~\cite{Wolfenstein:1977ue} give rise to  
MSW resonances~\cite{Mikheev:gs} when the matter density 
corresponds to
\beq
\rho_R = \pm \dmsq \cos 2\theta\;M_N / 
(2\sqrt{2}G_{\rm F} Y_e E)\; .
\label{msw}
\eeq 
Here $Y_e$ is the electron fraction, $M_N$ the average nucleon mass,
and the plus and minus signs 
correspond to neutrinos and antineutrinos respectively.
For neutrinos of energy $E$, resonances are possible at two 
densities, the $H$ resonance corresponding to 
$(\dmsq,\theta) \approx (\msqa,\theta_{13})$ and 
the $L$ resonance corresponding to 
$(\dmsq,\theta) \approx (\msqs,\theta_{12})$.
The energies of SN neutrinos are typically in the range
$5$--$50$ MeV.
For these energies, the $H$ resonance takes place around 
$\rho_H \sim 500$--$5000$ g/cc. It
occurs in neutrinos for normal hierarchy and 
in antineutrinos for inverted hierarchy.
The $L$ resonance that takes place around 
$\rho_L \sim 20$--$200$ g/cc always occurs in neutrinos. However, since $\theta_{12}$ is large, significant flavor conversions of antineutrinos also take place at the $L$ resonance.

The adiabaticities at the $H$ and $L$ resonances determine
the net neutrino conversion probabilities.
The $L$ resonance is always adiabatic, given the values
of $\msqs, \theta_{12}$ and the typical density profile of
the progenitor star around $\rho_L$.
The adiabaticity at the $H$ resonance is very sensitive to
the value of $\theta_{13}$ and the density profile of the
star in the resonance region. Indeed the neutrino conversion
rates are crucially dependent on the value of $\theta_{13}$,
and whether the $H$ resonance is in neutrinos or antineutrinos.
The SN signal can therefore be an extremely sensitive probe of
$\theta_{13}$ and whether the mass hierarchy is normal or 
inverted \cite{ds}.

In addition to divulging the neutrino mixing scenario,
SN neutrino fluxes can also allow us to have a peek at the
propagation of the shock wave while it is still inside the
mantle of the star.
The violent density fluctuations caused by the SN shock wave can
change the adiabaticity at the $H$ resonance in a time dependent manner, 
thus leaving their imprint on the time dependent neutrino spectra
\cite{fuller,takahashi,lunardini,fogli}.
In particular, the observations of the time dependent 
neutrino spectra can confirm the presence of forward as well as 
reverse shock wave through the ``double dip'' feature \cite{revshock}, 
and in addition can track the positions of the shocks
as they pass through the $H$ resonance region.
The feasibility of such a tracking at a water Cherenkov detector 
has been explored in \cite{fogli2,stochastic}.

Our understanding of the SN explosion mechanism is still
unsatisfactory \cite{janka,buras2}, which makes it very important to
extract as much information about the shock wave as possible.
In this paper, we demonstrate how the neutrinos that pass
through the shock wave near the $H$ resonance carry information
about the density profile of the shock wave.
In addition to making the $H$ resonance temporarily nonadiabatic,
the shock wave also forces the neutrinos to encounter multiple
$H$ resonances.
The relative phases that the neutrino mass eigenstates
gain between two or more of such $H$ resonances manifest
themselves as oscillations in the neutrino flavor
conversion probabilities as a function of the neutrino energy.
These oscillations are related to the
shock wave density profile, and in principle carry information about it. We therefore study the relation between the oscillations and the shock wave profile in detail.

Most of the analyses of SN neutrino conversions till now 
calculate the {\it probabilities} of conversions between 
neutrino mass eigenstates in matter in each resonance region,
and combine the results at different resonances, assuming them
to be independent. This includes the implicit assumption that
the information on the relative phase between the neutrino mass
eigenstates is lost between successive resonances. 
For neutrinos that undergo a single $H$ and a single $L$
resonance, this is a valid assumption since the two resonance
regions are well separated \cite{Keranen:2006gd}.
However, when the neutrinos encounter multiple $H$ resonances 
that are close to each other, coherence between
the neutrino mass eigenstates is maintained, and
one has to compute the {\it amplitudes} of neutrino flavor
conversions at the resonances, keeping track of the
relative phases. This gives rise to the phase effects that 
we explore in this paper. Such effects were pointed out 
in the context of solar neutrinos in \cite{Haxton:1986bc} 
and also observed in the numerical studies of supernova 
neutrinos in \cite{fogli2, Kneller:2005hf}. We provide an 
analytical approximation to study these effects, relating 
them to the the density profile of the medium. We also provide 
criteria to decide when decoherence or finite energy 
resolution of detectors renders these effects unimportant.

If the multiple resonances are semi-adiabatic, the phase
effects may be strong and if the oscillations are indeed observed
in the neutrino spectra, they can
help us infer about the density profile of the shock wave, which 
in turn can provide us important clues about the shock
wave propagation and the SN explosion.

The observability of these effects is dependent on factors like neutrino fluxes at the source, possible coherent development of neutrinos due to neutrino-neutrino interactions, the different density profiles encountered by neutrinos coming from different regions of the neutrinosphere, stochastic fluctuations of density, effective luminosity of the supernova, detector capabilities, et cetera. We comment on different model predictions of source fluxes and point out the model independent features of the phase effect. We investigate the number of events and detector capabilities required for the effect to be observable and find them to be demanding.

Our emphasis in this paper is to first ascertain the origin of these phase effects, their dependence on various parameters in the problem, and to check whether they are important at least in a simplified analysis where effects like coherent development of neutrinos \cite{coll-duan1, coll-raff}, anisotropy \cite{revshock} and turbulence \cite{stochastic, noise} of the medium are neglected. We find that even in a simplified setting, the observation of these effects is challenging.   

The paper is organised as follows.
In Sec.~\ref{analytic}, we give an analytical approximation for
calculating the neutrino conversion probability when multiple
$H$ resonances are taken into account in a two-neutrino framework.
The results of this section are general and can be applied
to any situation where multiple resonances are involved.
In Sec.~\ref{actual}, we apply the results to the
case of a SN shock wave, where the third neutrino and the 
$L$ resonance are included.
We show the neutrino conversion probabilities obtained by
using a realistic shock wave profile and 
study the feasibility of detecting the phase effects.
Sec.~\ref{concl} concludes.

\section{Phase effects from multiple resonances}
\label{analytic}

In this section, we calculate the survival probability of $\nu_e$
when they pass through multiple resonances, keeping track of
the relative phases between the mass eigenstates. 
The calculations are performed in the $2$-$\nu$ framework.
The results are readily extended to the $3$-$\nu$ framework in
the case of a SN shock wave, as will be shown in 
Sec.~\ref{actual}.
Although all the arguments in this section are given for
neutrinos, they are valid for antineutrinos just by changing the
sign of the matter potential $A(x)$. However, the flavor conversion analysis at the $H$ resonance is applicable to neutrinos in the normal mass hierarchy and for antineutrinos in the inverted mass hierarchy. Therefore, while analyzing antineutrinos, the value of $\dmsq$ should be taken to be negative.

\subsection{Survival probability of $\nue$ for a small mixing angle}
\label{anal}

The relevant mixing angle at the resonance $H$ is 
$\theta =\theta_{13}$,
on which we currently have a strong bound: 
$\sin^2 \theta_{13} < 0.05$ \cite{t13-bound}. 
Therefore, we try to solve
the problem using a small angle approximation.
We follow the notation and framework outlined in \cite{balantekin}
and work in the flavor basis.

Let $\nu_\beta$ be the relevant linear combination of 
$\nu_\mu$ and $\nu_\tau$.
When neutrinos pass through matter, their propagation is 
described by the Schr\"odinger equation
\beq
i \frac{d}{dx} \left( \begin{array}{c}
\nu_{e} \\ \nu_{\beta} \\
\end{array} \right)=H
\left( \begin{array}{c}
\nu_{e} \\ \nu_{\beta} \\
\end{array} \right).
\label{coupledeqns}
\eeq
Upto a matrix proportional to the unit matrix,the Hamiltonian $H$ is given by, 
\beq
\frac{1}{4E}\left( \begin{array}{cc}
A(x)- \dmsq \cos 2\theta & \dmsq \sin 2\theta \\
\dmsq \sin 2\theta & -A(x)+ \dmsq \cos 2\theta \\
\end{array} \right),
\label{hamiltonian}
\eeq
where $A(x) \equiv 2 E V(x) \equiv 2 \sqrt{2} G_F Y_e \rho E / M_N$.
These two coupled first order equations give rise to
the second order equation
\beq
-\frac{d^2}{dx^2}\nu_{e} - (\phi^2 + i\phi' )\nu_{e} 
= \eta^2 \nu_{e} \label{nu_eeqn}\\
\label{order2}
\eeq
where
\barr
\phi(x) = \frac{1}{4E}(A(x) - \dmsq\cos 2\theta )\; ,~~
\eta = \frac{\dmsq}{4E}\sin 2\theta
\earr
and prime ($'$) denotes derivative with respect to $x$.
In order to find the survival probability of $\nu_e$,
we solve for the $\nu_{e}$ wavefunction with the initial 
conditions $\nu_{e}(0) = 1$, $\nu_{\beta}(0) = 0$.
These conditions are equivalent to
\beq
\nu_{e}(0) = 1\; ,~~
i\left.\frac{d\nu_{e}}{dx}\right|_{0} = \phi(0)\; .
\label{initial}
\eeq
The ``logarithmic perturbation'' approximation
\cite{balantekin} solves 
the differential equation (\ref{nu_eeqn}) for
small mixing angles by choosing 
\beq
g \equiv 1 - \cos 2\theta
\label{g-def}
\eeq
as the small expansion parameter.
Denoting
\beq
\nu_{e} = e^{S(x)}\; , \quad  {\rm with} \quad  
S'(x) = c_0(x) + gc_1(x) + O(g^2)
\label{sx-def}
\eeq
so that
\barr
\nu_{e}(x) &=& \exp\left(\int_0^x{dx_1 c_0(x_1)}\right.\nonumber\\
& & \left.\phantom{\exp\left(\right.}+g\int_0^x{dx_1 c_1(x_1)}+ O(g^2)\right)\; ,
\earr
the solution becomes
\barr
\nu_{e}(x)  &=&  \exp\left[-\frac{iQ(x)}{2}
 - g\frac{i\dmsq x}{4E} \right. \nonumber\\
&& - g\frac{(\dmsq)^2}{2 ~(2 E)^2}\int_0^x dx_1 e^{iQ(x_1)}
\int_0^{x_1} dx_2 e^{-iQ(x_2)}\left. \phantom{\left(\right.\int\hspace{-0.65cm}}\right]\nonumber\\ 
  &&+ {\cal O}(g^2)~.
\label{og-soln}
\earr
Here we have defined the ``accumulated phase'' 
\beq
Q(x) \equiv \frac{1}{2 E}\int_0^x dx_1 \left( A(x_1)-\dmsq \right)
\; .
\label{Q-def}
\eeq
The survival probability $P_{ee}(x) \equiv 
P(\nu_{e}\rightarrow\nu_{e})$ at  $x=X$ then becomes
\barr
P_{ee}(X) &=& 
\exp\left[-g\frac{(\dmsq)^2}{2~(2E)^2}\left|
\int_0^X dx_1 e^{iQ(x_1)}\right|^2~\right]\nonumber\\ &&+ O(g^2)\; .
\label{Pee}
\earr
The integral in the above expression can be evaluated using the 
stationary phase approximation. 
The integral oscillates rapidly unless
$Q'(x) \approx 0$.
So the entire contribution to the integral can be taken to be
from the saddle point $x_s$, which
is the point where $Q'(x_s)=0$, i.e. 
\beq
A(x_s) = \dmsq \; .
\eeq
Note that this is also the resonance point in the small angle limit.

For a monotonic density profile, there is only one saddle point 
$x_s$ and the survival probability is
\beq
P_{ee} \approx \exp \left[-g \frac{\pi (\dmsq)^2}{2 E |A'(x_s)|}~\right]
\; ,
\label{pee-mono}
\eeq
which agrees with the Landau-Zener jump probability \cite{lz1,lz2}
in the limit of small mixing angle, and hence small $g$, even when
$P_{ee} \sim 1$. 

For a non-monotonic density profile, neutrinos can experience
more than one resonance at the same density but at different positions. 
In that case $Q'(x) =0$ at more than one point.
If the resonances are sufficiently far apart,
the contributions from each of them  may be added 
independently of each other. Their total contribution 
to the integral in (\ref{Pee}) is
\beq
\int_0^X dx e^{iQ(x)} \approx \sum_ie^{i\alpha_i}e^{iQ(x_{i})}
\left({\frac{4\pi E} {|A'(x_{i})|}} \right)^ {{1/2}}
\eeq
where $i$ runs over all the saddle points. 
Note that $\alpha = \pi /4$ if $A'(x_{s}) < 0$ and  
$\alpha = 3\pi /4$ if $A'(x_{s}) > 0$. 
The probability calculated using (\ref{Pee}) now has 
terms which depend on the differences between the integrated phases
\barr
\Phi_{ij} & \equiv & Q(x_{j})-Q(x_{i})+\alpha_j-\alpha_i \nonumber\\ & = & 
\int_{x_{i}}^{x_{j}} \frac{1}{2E}\left(A(x)-\dmsq \right) dx + \alpha_{j}-\alpha_{i}\; .
\label{phi-ij-def}
\earr
In general,
\barr
P_{ee}(X) &=& \exp \left[ -g
\left( \sum_i a_i^2 + 2 \sum_{i<j} a_i a_j \cos \Phi_{ij} \right)
\right]\nonumber\\ &&+ {\cal O}(g^2)
\label{Pee-general}
\earr
where
\beq
a_i \equiv \left( \frac{\pi (\dmsq)^2}{2 E |A'(x_{i})|} \right) ^{1/2}
\; .
\label{a-def}
\eeq

For example, when there are only two saddle points the survival 
probability is given by
\barr
P_{ee} &=& \exp(-g a_1^2) \exp(-g a_2^2) \exp(-2 g a_1 a_2 \cos\Phi_{12}) \nonumber\\
       & &+ {\cal O}(g^2)\; . 
\label{master-eqn}
\earr
The first two factors in (\ref{master-eqn})
are the individual Landau-Zener jump
probabilities for the two level crossings.
The last factor gives rise to oscillations in $P_{ee}$ as a function of energy.
The oscillation pattern has its maxima at
$\Phi_{12} = (2 n + 1) \pi$ 
and minima at  
$\Phi_{12} = 2 n \pi$
where $n$ is an integer. This expression is valid as long as $g a_{i}^2 \lsim 1$, so that the ${\cal O}(g^2)$ terms remain small, and none of the resonances overlap.

\begin{figure}
\begin{center}
\epsfig{file=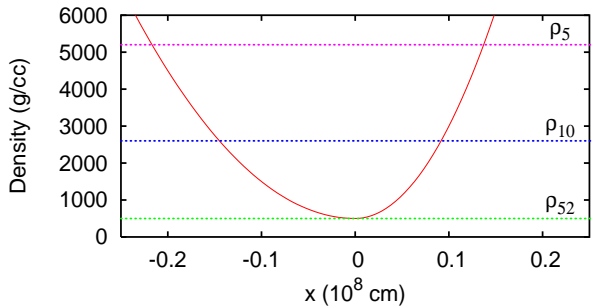,width=3.3in,height=1.65in,angle=0}
\caption{Density profile in (\ref{toy-ne}) with $a=500$ g/cc, 
$b_1 = 10^5 \mbox{ g/cc } / (10^8 \mbox{ cm})^2$ and 
$b_2 = 2.5 \times 10^5 \mbox{ g/cc } / (10^8 \mbox{ cm})^2$ . The horizontal lines on the graph are the resonance densities for various energies ($5$, $10$, $52$ MeV) taking $\theta = 0.02$ rad $\approx 1.1^\circ$. 
Notice how the saddle points come closer for larger energies till 
$E=E_{R(max)}=52$ MeV, after which there is no resonance. 
\label{densityplot}}
\end{center}
\end{figure}

We illustrate the validity and limitations of the 
small angle approximation with a toy density profile
\beq
\rho(x) = \left\{
\begin{array}{ll}
a + b_1 x^2 & (x<0) \\
a + b_2 x^2 & (x>0) \\
\end{array}
\right. \; ,
\label{toy-ne}
\eeq
as shown in figure~\ref{densityplot}. We take $Y_e=0.5$ and 
$\dmsq=0.002~{\rm eV}^2$.
Neutrinos are produced at $x \to -\infty$ and we calculate
$P_{ee}$ at $x \to \infty$.
We also show the positions of resonance densities for various energies, which are given by
\beq
\rho_R[{\rm g/cc}]\approx \pm \frac{\dmsq [{\rm eV}^2]\cos 2\theta}{2\times7.6\times10^{-8} Y_e E[{\rm MeV}]}~.
\label{msw2}
\eeq

\begin{figure}
\begin{center}
\epsfig{file=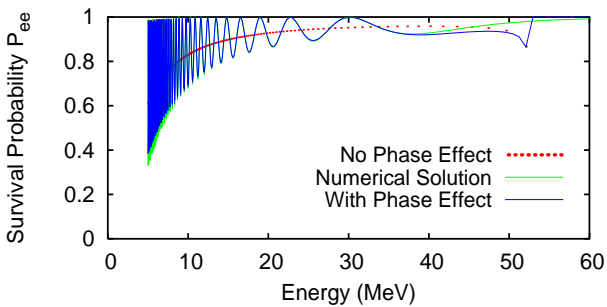,width=3.3in,height=1.65in,angle=0}
\caption{Survival probability $P_{ee}$ as a function of energy for
$\theta =0.02$ rad $\approx 1.1^\circ$. 
The green (light) curve is the numerically evaluated exact result.
The blue (dark) curve is our solution with small angle approximation
including the phase effects. 
The red (dotted) curve is the approximate solution 
if the phase effects are neglected.
Notice that our approximate solution is valid only upto $E=E_{R(max)}$.}
\label{nonmonoplot}
\end{center}
\end{figure}

Figure \ref{nonmonoplot} shows the survival probability $P_{ee}$
as a function of energy, both the exact numerical result and the
result of our analytical approximation for small angles. 
It can be seen that at such small angles ($\theta = 0.02$ rad
$\approx 1.1^\circ$), the approximation works extremely well.

Note that the amplitude of the oscillations
is comparable to the deviation of the average survival probability
from unity. That is, the oscillation effect is not a small effect. 
Indeed, the oscillation term is of the same order as the averaged
effect, as can be seen from (\ref{master-eqn}).
Figure~\ref{nonmonoplot} also shows the average value of
$P_{ee}$ that one would have obtained if one 
naively combined the jump probabilities at the two resonances.
Our analysis gives additional 
oscillations in the survival probability as a function of 
neutrino energy about this average value.
This effect is what we call as the phase effect, 
and is clearly significant as can be seen from the figure.
An important feature of the oscillations is that 
the ``wavelengths,'' i.e. the distances between the consecutive
maxima or minima,  are larger at larger $E$.

\begin{figure}
\begin{center}
\epsfig{file=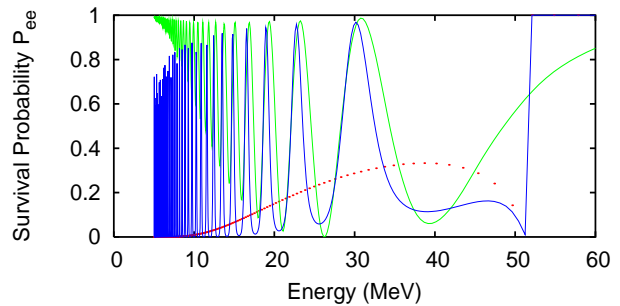,width=3.3in,height=1.65in}
\caption{Survival probability $P_{ee}$ as a function of energy for
$\theta =0.1$ rad $\approx 5.7^\circ$. 
The convention for the lines is the same as that used in 
figure~\ref{nonmonoplot}.
\label{pee-large}}
\end{center}
\end{figure}

The resonances start overlapping at $E \approx E_{R(max)}$
(figure~\ref{densityplot}), 
which is where our approximation starts breaking down,
as can be seen in figure~\ref{nonmonoplot}.
For $E > E_{R(max)}$, the neutrinos no longer encounter a strict
resonance, and our approximation gives $P_{ee}=0$ identically.
However, the resonances have finite widths which may affect the
conversion probabilities of neutrinos with $E \approx E_{R(max)}$.
The sharp jump observed in figure~\ref{nonmonoplot} 
at $E \approx 52$ MeV is therefore
not a real effect, but a limitation of our technique.

The small angle approximation starts failing for larger angles and
lower energies, where $ga_{i}^2 \gsim {\cal O}(1)$.  Figure~\ref{pee-large} shows that the amplitude
at low energies is not calculated correctly for $\theta = 0.1$ rad 
$\approx 5.7^\circ$. However, note that the positions of maxima and minima 
of $P_{ee}$ are still predicted to a good accuracy. We shall
argue in the next subsection that these 
can be computed accurately for the
whole allowed range of $\theta_{13}$,
given the non-monotonic density profile between the
two resonances.

\subsection{The oscillations in $P_{ee}(E)$}
\label{freq}

Let us consider a density profile with a ``dip'' as in the
toy model in the previous section.
A neutrino with energy $E$ encounters two
resonances $R_1$ and $R_2$ at $x=x_1$ and $x=x_2$ respectively,
so that
\beq
\rho_R \equiv \rho(x_1) = \rho(x_2) \; .
\eeq
We assume $Y_e$ to be a constant throughout the region of interest.
We also assume that the propagation of neutrino mass eigenstates is 
adiabatic everywhere except in the resonance regions
$(x_{1-}, x_{1+})$ and $(x_{2-},x_{2+})$ around the resonance
points $x_1$ and $x_2$ respectively.
In the limit of a small mixing angle, the widths of the resonances are
small:
\beq
\Delta \rho \approx \rho \tan 2\theta \; .
\eeq
Therefore, $x_{1-}\approx x_1 \approx x_{1+}$ and
$x_{2-}\approx x_2 \approx x_{2+}$. We shall work in this
approximation, and shall use the notation $x_{i\pm}$ only
for the sake of clarity wherever needed.

At $x \ll x_1$, the density $\rho(x) \gg \rho_R$,
so that the heavier mass eigenstate $\nu_H$ is approximately
equal to the flavor eigenstate $\nu_e$. Let us start with $\nu_e$
as the initial state:
\beq
\nu_e (x \ll x_1) \approx \nu_H \; . 
\eeq
The mass eigenstate $\nu_H$ propagates adiabatically till it
reaches the resonance region $x \approx x_1$: 
\beq
\nu_e (x_{1-}) \approx \nu_H \; .
\eeq
While passing through the resonance, unless the
resonance is completely adiabatic, the state becomes 
a linear combination of $\nu_H$ and $\nu_L$, the lighter
mass eigenstate. Note that the phases of $\nu_H$ and $\nu_L$
can be defined to make their relative phase vanish at
$x=x_{1+}$.
\beq
\nu_e(x_{1+}) = \cos\chi_1 \nu_H + \sin\chi_1 \nu_L \; ,
\eeq
where $P_1 \equiv \sin^2 \chi_1$ is the ``jump probability''
at $R_1$ if it were an isolated resonance.

The two mass eigenstates $\nu_H$ and $\nu_L$ propagate to the
other resonance $R_2$, gaining a relative phase in the process
(the overall phase of the state is irrelevant):
\beq
\nu_e(x_{2-}) = \cos\chi_1 \nu_H 
+\sin\chi_1 \exp \left( i \int_{x_{1}}^{x_{2}} 
\frac{\Delta \tilde m^2}{2E} dx 
\right) \nu_L \; ,
\eeq
where $\Delta \tilde m^2$ is the mass squared difference
between $\nu_H$ and $\nu_L$ in matter:
\beq
\Delta \tilde m^2 = \left((\Delta m^2 \cos 2\theta - 2 E V(x))^2  +  (\dmsq \sin 2\theta)\right)^{1/2}~.
\label{dmtilde}
\eeq

The effect of the resonance $R_2$ may be parametrised in general as
\beq
\left(\hspace{-0.1cm} \begin{array}{c}
\nu_H(x_{2+}) \\ \nu_L(x_{2+}) \\ \end{array} \hspace{-0.1cm}\right) =
\left( \begin{array}{cc}
\cos\chi_2 & \sin \chi_2 e^{i\varphi} \\
-\sin\chi_2 e^{-i\varphi} & \cos\chi_2 \\ \end{array} \right) 
\left(\hspace{-0.1cm} \begin{array}{c}
\nu_H(x_{2-}) \\ \nu_L(x_{2-}) \\ \end{array}\hspace{-0.1cm} \right). 
\eeq
where $P_2 \equiv \sin^2 \chi_2$ is the ``jump probability''
at $R_2$ if it were an isolated resonance.

From (\ref{og-soln}), one can deduce that in the limit
$x_{2-} \approx x_{2+}$, we have $\varphi \approx Q(x_{2+}-x_{2-})
\approx 0$. 
The state $\nue(x_{2+})$ can then be written as
\barr
\nu_e(x_{2+}) &=&  
\left[\phantom{\int}\right.\hspace{-0.35cm}\cos\chi_2 \cos\chi_1 \nonumber \\ &&+ \sin \chi_2\sin\chi_1 
\exp \left( i \int_{x_{1}}^{x_{2}} \frac{\Delta \tilde m^2}{2E} dx \right) \left.\phantom{\int}\hspace{-0.35cm}\right] \nu_H
\nonumber \\
&&+  \left[\phantom{\int}\right.\hspace{-0.35cm} \cos\chi_2 \sin\chi_1 
\exp \left( i \int_{x_{1}}^{x_{2}} \frac{\Delta \tilde m^2}{2E} dx \right) \nonumber \\
&&-\sin\chi_2 \cos\chi_1 \left.\phantom{\int}\hspace{-0.35cm}\right] \nu_L\; .
\earr

For $x > x_{2+}$, the mass eigenstates travel independently 
and over sufficiently large distances, decohere from
one another. At $x \gg x_2$, since $\rho(x) \gg \rho_R$, the
heavier mass eigenstate $\nu_H$ again coincides with $\nu_e$
and we get the $\nu_e$ survival probability as
\barr
P_{ee} & = & 
\cos^2(\chi_1-\chi_2)\nonumber \\&&- \sin 2\chi_1 \sin 2\chi_2
\sin^2 \left( \int_{x_{1}}^{x_{2}} 
\frac{\Delta \tilde m^2}{4E} dx \right)\; .
\label{eq-freq}
\earr

If the phase information were lost, either due to decoherence or
due to finite energy resolution of the detectors \cite{decoherence}
the survival probability would have been
\barr
\hspace{-0.3cm}P_{ee \mbox{\footnotesize (no phase)}} &=& P_1 P_2 + (1-P_1) (1-P_2)\nonumber\\
&=& \cos^2 \chi_1 \cos^2 \chi_2 + \sin^2 \chi_1 \sin^2 \chi_2\;,
\label{prob-combo}
\earr
which matches with (\ref{eq-freq}) when the $\sin^2(\int ..)$ term
is averaged out to $1/2$.

The $\sin^2(\int ..)$ term in (\ref{eq-freq}) gives rise
to the oscillations in $P_{ee}(E)$.
If two consecutive maxima of $P_{ee}$ are at energies $E_k$ 
and $E_{k+1}$ such that $E_k > E_{k+1}$, then the condition
\barr
\int_{x_1(E_{k+1})}^{x_2(E_{k+1})} 
\frac{\Delta \tilde m^2 (x, E_{k+1})}{2 E_{k+1}} ~dx &-&
\int_{x_1(E_k)}^{x_2(E_k)} 
\frac{\Delta \tilde m^2 (x, E_k)}{2 E_k} ~dx  
\nonumber\\ &=& 2 \pi
\label{integrals}
\earr
is satisfied. The quantity $(E_k - E_{k+1})$ is the ``wavelength''
of the oscillations.

Note that $\Delta \tilde m^2(x,E)$
is equal to $|A(x) -\Delta m^2|$ in the small angle limit.
Moreover, this quantity is rather insensitive to $\theta$ 
in the allowed range of $\theta_{13}$. 
Therefore, 
it is not a surprise that the predictions of 
the positions of maxima and minima 
in the small angle approximation (Sec.~\ref{anal}) 
are accurate and robust in the whole range $\theta < 13^\circ$.

Since $\theta$ is small, the left hand side of (\ref{integrals})
is approximately equal to 
the area of the region in the density profile plot
enclosed by the densities 
$\rho_{E_k}$ and $\rho_{E_{k+1}}$:  
\beq
{\cal A} \approx 2 \pi \frac{M_N}{\sqrt{2} G_F Y_e} \; .
\eeq
The distance between the two resonances in 
the region $\rho_{E_k} < \rho < \rho_{E_{k+1}}$ is then 
\beq
 r_k \approx  {\cal A} / (\rho_{E_{k+1}} - \rho_{E_{k}})\; .
\label{res-dis}
\eeq
This procedure may be repeated for various $k$ values
to estimate the separation between the resonances at the 
corresponding densities, and hence to constrain the form of the
density profile. Although this seems straightforward in our simplified analysis, the effect of density variations due to convection, turbulence and anisotropies would greatly complicate such a reconstruction in practice.

Nonmonotonic density profiles are encountered by neutrinos
escaping from a core collapse supernova during the shock wave
propagation. If the phase effects are observable at neutrino
detectors, the above procedure may help us reconstruct the
shock wave partially. Of course the procedure is somewhat 
crude, and the information obtained would only be on the density 
profile along the line of sight.
We have also assumed that neutrinos coming from different
parts of the neutrinosphere encounter nearly the same
density profiles.
However, since neutrinos are the only particles that
can even in principle carry information from so deep inside
the exploding star, it is important to check whether
the detection of these phase effects is feasible even with these simplifying assumptions. We shall do this in the next section.

\section{Oscillations during the SN shock wave propagation}
\label{actual}

In this section, we apply the results in the last section to the
neutrinos travelling through a supernova shock wave. Though
we have to consider three-neutrino mixing in this case, the separation
of $H$ and $L$ resonances \cite{kuo} means that we can calculate
the transition probabilities at these resonances separately.
Each of these resonances can then be treated as an effective
two-neutrino level crossing. 
The $L$ resonance that takes place
in neutrinos is always adiabatic \cite{ds}. If the net survival probability after passing through all the $H$ resonances is denoted by $P_H$, the survival probability of $\nue$ after passing through all the $H$ and $L$ resonances is
\beq
p = P_H \sin^2 \theta_\odot \quad {\rm (NH) , } \quad
p = \sin^2 \theta_\odot \quad {\rm (IH) } 
\label{p-def}
\eeq
where NH and IH stand for normal and inverted mass hierarchy 
respectively. 
Here $\theta_\odot$ is the solar mixing angle.
Similarly, the survival probability of $\nuebar$ after passing 
through all the $H$ and $L$ resonances is
\beq
\bar{p} = P_H \cos^2 \theta_\odot \quad {\rm (IH) , } \quad
\bar{p} = \cos^2 \theta_\odot \quad {\rm (NH) } \; .
\label{pbar-def}
\eeq
Clearly, since the phase effects appear through $P_H$,
they will be visible only in $\nu_e$ for normal hierarchy and
only in $\nuebar$ for inverted hierarchy.

In present and planned water Cherenkov \cite{hk}
and scintillation \cite{lena} detectors, 
the main  neutrino detection channel is the inverse beta decay
reaction $\bar\nu_e p\to ne^+$ that allows the reconstruction
of $\nuebar$ energies. Therefore we consider only the $\bar\nu_e$
spectrum in our analysis.
However an analogous analysis can be easily performed in the neutrino
channel for a detector able to measure the $\nue$ spectrum, 
for example using liquid argon \cite{liq-ar}.

\begin{figure}
\begin{center}
\epsfig{file=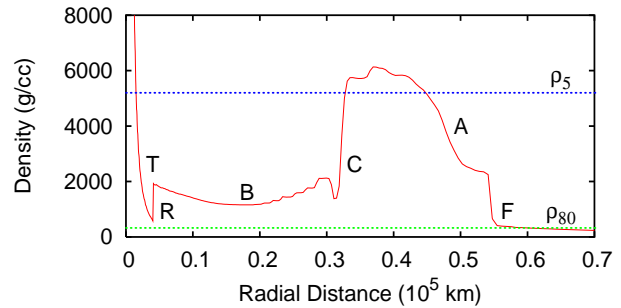,width=3.3in,height=1.65in}
\caption{Snapshot of a shock wave density profile at $t=5$ sec.
The resonance densities for $E=5$ and $80$ MeV (with $Y_e=0.5$)
are also shown. We take $\dmsq=0.002~{\rm eV}^2$ and 
$\theta = 0.02$ rad $\approx 1.1^\circ$.
\label{snapshot}}
\end{center}
\end{figure}

In order to illustrate the phase effects on $p$ or $\bar{p}$,
we consider a typical snapshot of the density profile of a SN
during a shock wave \cite{revshock}, as shown in 
figure~\ref{snapshot}. 
The forward shock F and the reverse shock R are 
sharp density discontinuities, the density change 
of a factor of two or more taking place
over a distance of much less than a km.
The density variation in the ``contact discontinuity'' C,
which is the transition region between the shock-accelerated
and neutrino-heated SN ejecta, takes place more slowly, over a
distance of more than $100$ km \cite{revshock}.
The mass accretion region A behind the forward shock wave, and 
the low density bubble B have gradually changing densities.
The region T is the tail of the shock wave.

The neutrinos, while passing through these regions, may undergo 
multiple level crossings.
The extent of flavor conversion in each region will depend on the value
of $\theta_{13}$ and the steepness of the density profile in
that region.
It is found that for $\theta \sim 0.01$ rad or higher, 
the density variations in the mass accretion region A, the 
low density bubble B and the contact discontinuity C are
too gradual for any non-adiabaticity.
We therefore concentrate on the forward shock F, the reverse shock
R and the tail T. 

The coherence between mass eigenstates and the oscillations in the survival probability may be lost due to two major sources: the separation of mass eigenstate wavepackets and the finite energy resolution of the detector. The coherence length $L_{coh}$, defined as the distance over which the wavepackets separate, is given by~\cite{Nussinov} 
\beq
L_{coh} \sim 4\sqrt{2} \sigma E^{2}/ \dmsq ~,
\label{coherence-crit}
\eeq
where $\sigma$ is the width of the wavepacket at source. Taking $\sigma \sim 10^{-9}$ cm near the neutrinosphere ~\cite{Anada:1989fk} in the relevant energy range of $5$--$80$ MeV, the coherence length for SN neutrinos is $L_{coh} \sim 10^{3}$--$10^{5}$ km. Resonances separated by distances larger than $L_{coh}$ may be taken to be incoherent. Since the distances involved are ${\cal O}(10^4~{\rm km})$ (See figure~\ref{snapshot}), a definite conclusion about decoherence due to wavepacket separation cannot be reached with this simple estimate. However, for observability the oscillations in survival probability must also occur over energy intervals much larger than $\delta E_{det}$, the uncertainity in energy measurement at the detector. This turns out to be the dominant factor in smearing out the oscillations. We can estimate from (\ref{integrals}) the difference in energies at which successive maxima of the survival probability occur to be
\beq
\lambda_{E_{k}} \equiv E_{k+1}-E_{k} \sim 4 \pi E_{k}^2/ \dmsq r_{k}\;,
\label{ener-crit}
\eeq
where $r_{k}$ is the distance between the two resonances encountered by a neutrino of energy $E_{k}$. For the energy range of $5$--$80$ MeV, this gives $\lambda_{E_{k}}$ of $1$--$10$ MeV for the T-R resonance pair with $r_{k}\sim10^3$--$10^4$ km. The oscillations are faster for the  T-F and R-F pairs that occur about $10^5$--$10^6$ km apart. Typically the energy resolution is $1$--$10$ MeV for a water Cherenkov detector and $0.1$--$1$ MeV for a scintillation detector over this energy range. Moreover, the charged lepton energy is not of the same as the energy of the neutrino, which introduces additional smearing. Indeed as we shall see, the fast oscillations due to the T-F or R-F resonance pairs will be smeared and therefore clearly unobservable even at a scintillation detector. 
Note that the above arguments are only qualitative; 
we have neglected the matter effects while estimating $\L_{coh}$ and 
$\lambda_{E_k}$.

\begin{figure}
\begin{center}
\epsfig{file=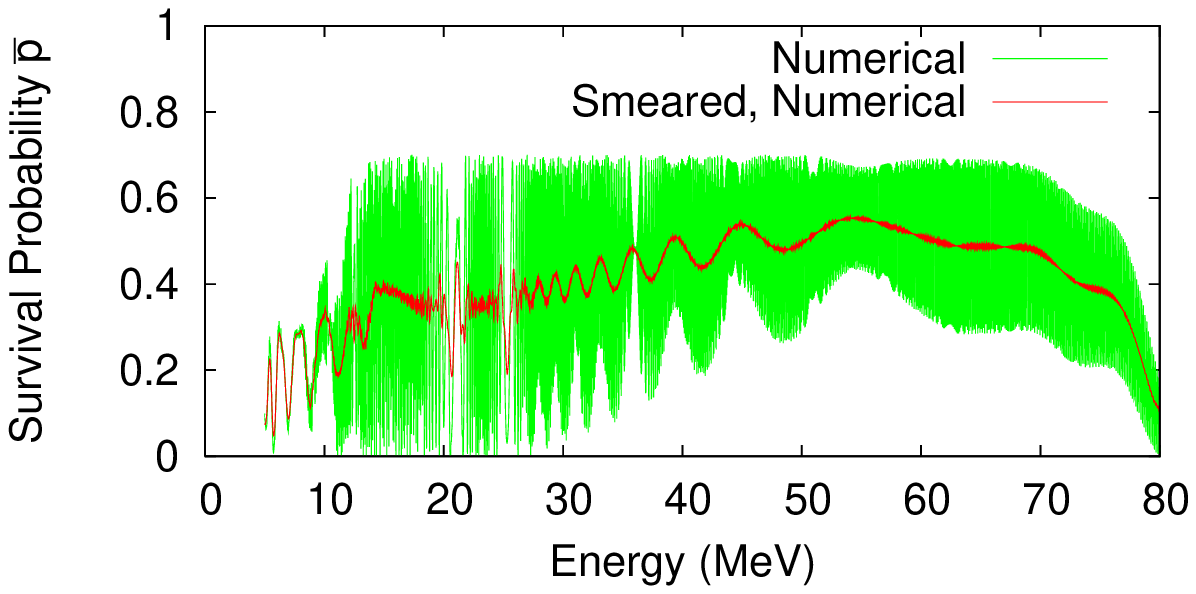,width=3.3in,height=1.65in}
\caption{Survival probability $\bar{p}$ of $\nuebar$
for inverted hierarchy with the density profile in
figure~\ref{snapshot}.
The smeared probability is obtained by taking
a running average over the energy range corresponding to
the typical energy resolution of a scintillation detector i.e., $\Delta E_{\rm SC} {\rm (MeV)}  \approx 0.2 \sqrt{E/10~ {\rm MeV}}$.
\label{pbarplot}}
\epsfig{file=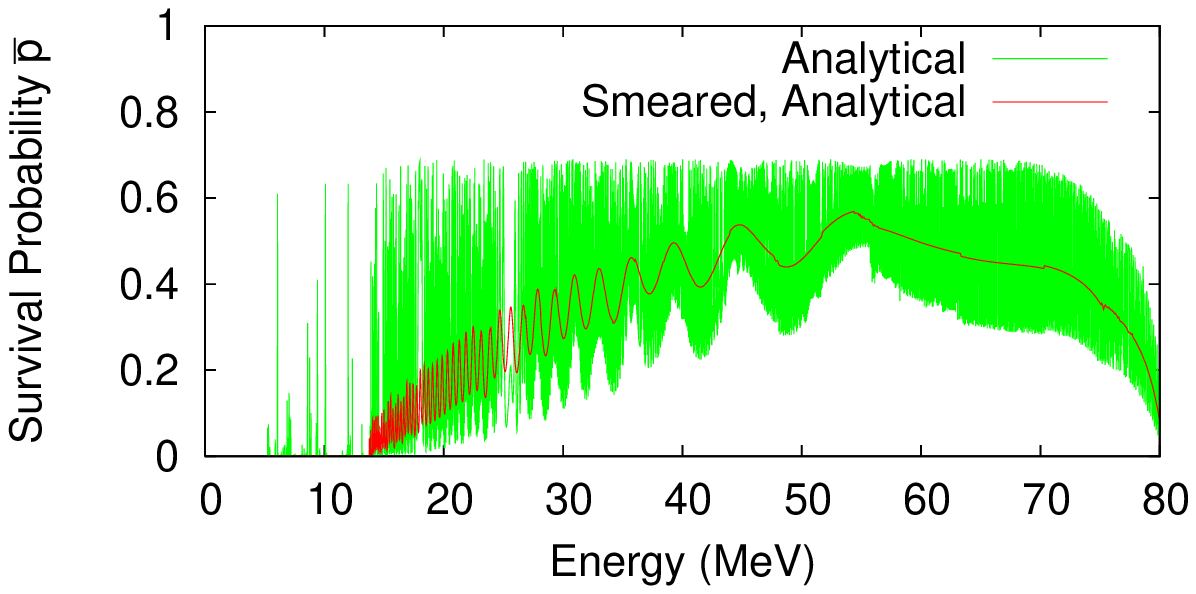,width=3.3in,height=1.65in}
\caption{Same as figure \ref{pbarplot}, calculated analytically using (\ref{Pee-general}) for resonances at T, R, F. The smeared curve is obtained by dropping the fast oscillatory terms. Note that analytical approximation reproduces the numerical result reasonably well for $E>20$ MeV. 
\label{analpbarplot}}
\epsfig{file=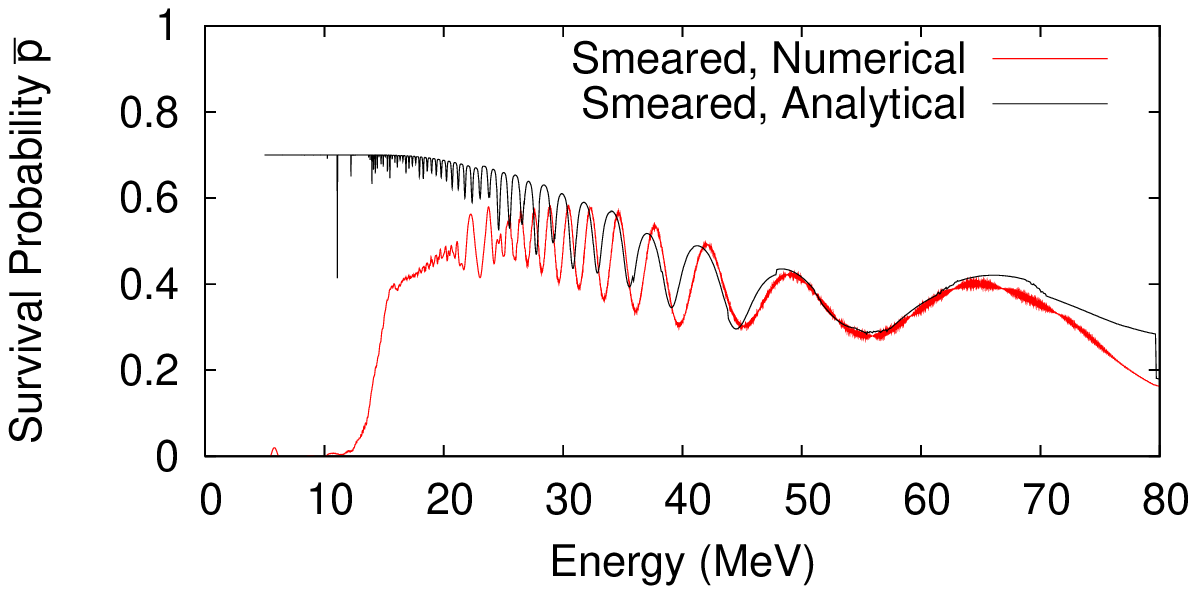,width=3.3in,height=1.65in}
\caption{Comparison of smeared probability calculated numerically and using the analytical approximation, for a larger mixing angle $\theta_{13}=0.05$ rad $\approx 2.9^\circ$. Note that analytical approximation reproduces the numerical result reasonably well for $E>25$ MeV.
\label{pbarlargetheta}}
\end{center}
\end{figure}

Figure~\ref{pbarplot} shows the value of $\bar{p}$ as a 
function of energy for $\theta=0.02$ rad $\approx 1.1^\circ$.
The rapid oscillations correspond to the relative phase
$\Phi_{RF} (\approx \Phi_{TF})$ that is accumulated
by the mass eigenstates between resonance regions R and F
(T and F). Such high frequency oscillations are virtually
impossible to observe, given the practical limits on the energy
resolutions of neutrino detectors.

We smear the high frequency oscillations by taking
a ``running average'' over the energy range corresponding to
the typical energy resolution of a scintillation detector.
The low frequency oscillations that survive 
are found to correspond to the relative phase $\Phi_{TR}$
accumulated by the mass eigenstates between resonances in 
regions T and R. 
Since these two resonances are closer compared to the resonance pairs
R-F or T-F, the frequency of oscillations is smaller.
The same oscillation pattern is observed if the survival probability
is computed by assuming that the resonance in region F is 
completely non-adiabatic,
which confirms that the pattern is indeed due to the  
level crossings in regions T and R.


Note that $\bar{p}$ goes to $\cos^2\theta_\odot$ at its maximum where $P_H$ goes to unity, whereas at $\bar{p}~\approx 0$ at high energies and low energies where $P_H$ goes to zero. 
The oscillations in the low energy region ($E<~20$ MeV) are 
too rapid to be observable. The fluctuations observed in the
running average at low energies are not robust: they depend
partly on the details of the density profile and are partly 
numerical artifacts. 

In figure~\ref{analpbarplot}, we plot the value of $\bar{p}$ for the same parameter values, but calculated using (\ref{Pee-general}) at the T, R, F resonances. The smeared probability curve is calculated by dropping the oscillatory terms due to the resonance at F and combining the survival probability at the T-R pair with that at F using (\ref{prob-combo}). We see that the analytical expression agrees quite well with the numerical result of figure \ref{pbarplot}. In particular, the oscillation frequency matches quite well. 
Moreover, the slope of $A(x)$ at T is about $0.1$ times that at R, 
therefore (\ref{a-def}) and (\ref{master-eqn})
predict the amplitude of oscillations 
to be $\sqrt {A'(T)/A'(R)}~\approx 0.3$ times the mean 
value of $\bar{p}$. 
This estimate is also in good agreement with our numerical results
in the figure.

The analytical approximation in (\ref{Pee-general}) breaks down when $g a_{i}^2 > {\cal O}(1)$. This happens below a certain value of energy, that is higher for larger mixing angles. Comparing figure~\ref{pbarplot} and figure~\ref{analpbarplot} we see that for $\theta=0.01$ rad the approximation is resonably accurate for neutrino energies above $20$ MeV. Figure~\ref{pbarlargetheta} shows the smeared $\bar{p}$ calculated numerically and analytically for $\theta=0.05$ rad. It is clear that the approximation is valid above $25$ MeV. The approximation fails at low energies where $\bar{p}$ calculated analytically does not tend to zero. For higher energies, the oscillations due to the phase effect are predicted quite accurately.

\begin{figure}
\begin{center}
\epsfig{file=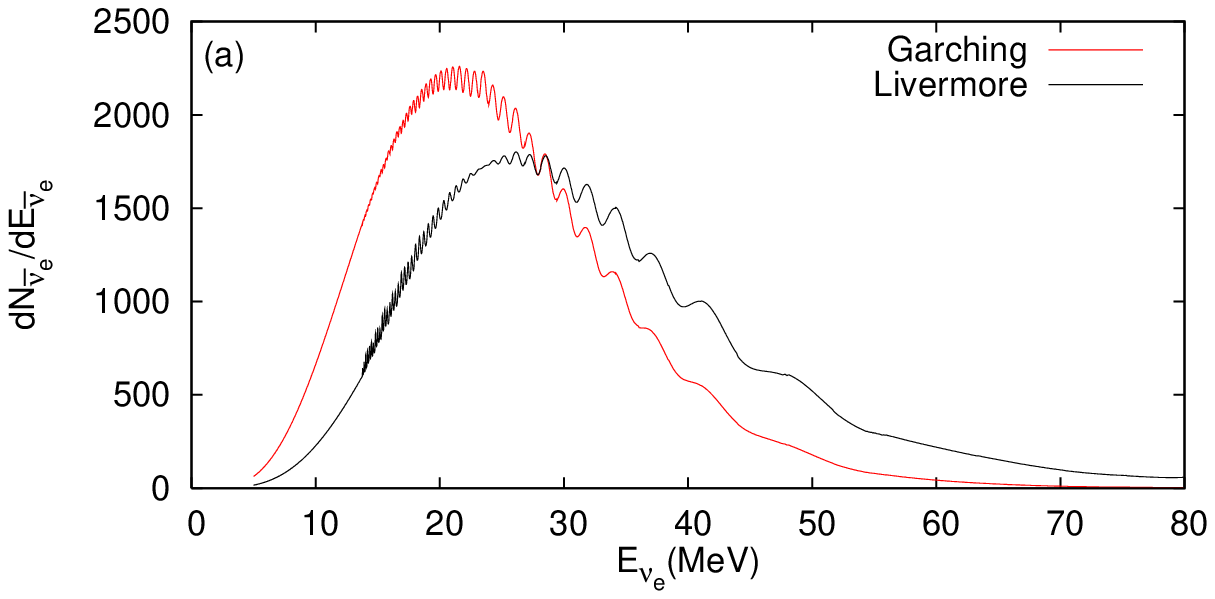,width=3.3in,height=1.65in}
\epsfig{file=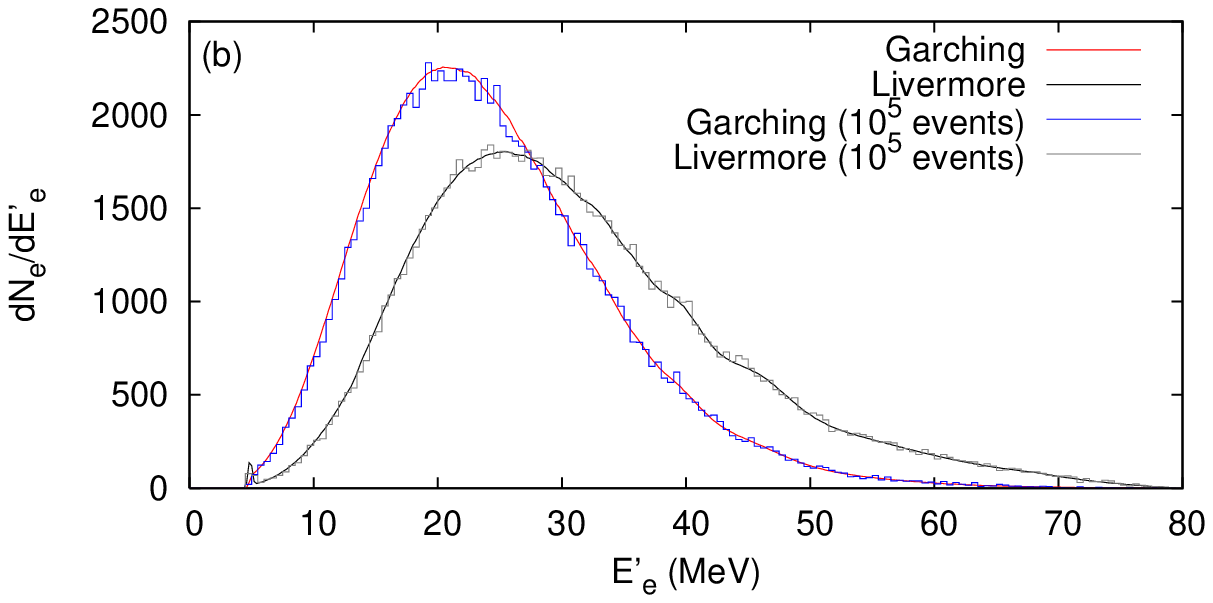,width=3.3in,height=1.65in}
\caption{
The upper figure (a) shows the oscillations in the energy 
spectra of the $\nuebar$,
if the $\nuebar$ energy were accurately measurable,
for the Garching and Livermore models.
The bottom figure (b) shows the observed positron energy
spectra in a scintillation detector.
All the spectra are normalized so that 
the total number of events is $10^5$.
The figure (b) also shows the statistical fluctuations for
$10^5$ events with an energy binning of $0.5$ MeV.
\label{fig-flux}}
\end{center}
\end{figure}

In order to explore the observability of the phase effects,
we use the parametrization for the primary fluxes given by
\cite{Keil:2002in}:
\beq
F^0_{\nu_i}=
\frac{\Phi_0}{E_0}\,N(\alpha)\left(\frac{E}{E_0}\right)^\alpha
\exp\left[-(\alpha+1)\frac{E}{E_0}\right] \; ,
\eeq
where $N(\alpha)~=~(1+\alpha)^{1+\alpha}/\Gamma(1+\alpha)$.
For illustration, we choose two models of neutrino fluxes, the 
Garching model \cite{garching} that uses the parameters
\barr
\alpha_{\nuebar} = \alpha_{\bar{\nu}_x} =3 \; , ~
E_0(\nuebar) = 15~ \mbox{\rm MeV} \;, \nonumber\\   
E_0(\bar{\nu}_x) = 18~ \mbox{\rm MeV} \;,
~\Phi_0(\nuebar)/\Phi_0(\bar{\nu}_x) = 0.8 \; ,\nonumber
\earr
and the Livermore model \cite{livermore} that uses
\barr
\alpha_{\nuebar} = \alpha_{\bar{\nu}_x} =3 \;, ~
E_0(\nuebar) = 15~\mbox{\rm MeV} \; ,\nonumber\\ ~  
E_0(\bar{\nu}_x) = 24~\mbox{\rm MeV} \; ,~
\Phi_0(\nuebar)/\Phi_0(\bar{\nu}_x) = 1.6\; .
\earr
We plot in figure~\ref{fig-flux}(a) the quantity
\barr
dN_\nuebar/dE_\nuebar & = &
\sigma(E_\nuebar) F_{\nuebar}\nonumber\\ & = & 
\sigma(E_\nuebar) [F^0_{\bar{\nu}_x} + 
\bar{p} (F^0_{\nuebar}-F^0_{\bar{\nu}_x})]
\label{sigma-e}
\earr
as a function of the neutrino energy $E_\nuebar$,
where we have normalized the spectrum such that the total 
number of events is $10^5$.
Here we use the differential cross section as computed in 
\cite{strumia}.

The phase effect is most prominent around $4$--$5$ seconds postbounce when the tail T and reverse shock R pass through the $H$ resonance and the T-R resonance pair is about $10^3$ km apart. We expect from (\ref{ener-crit}) that the extrema in the survival probability will occur at energies separated by about $1$--$10$ MeV. This is clearly visible in figure~\ref{fig-flux}(a). Moreover, the positions of the extrema are independent of the primary fluxes.

However, the spectrum shown in figure~\ref{fig-flux}(a) 
is not directly observable: one can only observe the energy 
spectrum of positrons produced by the inverse beta reaction. 
Assuming quasielastic scattering, the positron energy is 
given by \cite{strumia}
\beq
E_e=\frac{(E_\nu-\delta)(1+\epsilon)
+ \epsilon\cos\vartheta
\sqrt{(E_\nu-\delta)^2-m_e^2 \kappa}}{\kappa}
\label{E-positron}
\eeq 
and 
\beq
p_e=\sqrt{E_e^2-m_e^2}\; ,
\label{e-positron}
\eeq
where $\vartheta$ is the angle of scattering, 
$M_p$ the proton mass,
$\epsilon \equiv E_\nuebar/M_p$, and
$\kappa \equiv (1+\epsilon)^2 - (\epsilon \cos\vartheta)^2$.
For $E \approx 40$ MeV, we have $\epsilon \approx 1/25$, so that
the positron energy is spread over a range of $\approx 4$ MeV
depending on the scattering angle.
Given that the successive maxima of the oscillation 
pattern are separated by only about $2$--$8$ MeV in this energy range,
the oscillation pattern is significantly smeared out.

A further smearing of the oscillation pattern is caused by 
the finite energy resolution of the detector.
The energy resolution of a water Cherenkov detector is typically
$\Delta E_{\rm CH}{\rm (MeV)}  \approx 1.6 \sqrt{E/10~{\rm MeV}}$
and washes off the oscillations completely.
For a scintillation detector, the resolution is much better,
$\Delta E_{\rm SC} {\rm (MeV)}  \approx 0.2 \sqrt{E/10~{\rm MeV}}$.
We show in figure~\ref{fig-flux}(b) the spectrum
of the observed positron energy $E'_e$
after taking (\ref{E-positron}) and (\ref{e-positron}) into account 
and using the energy 
resolution of a scintillation detector.
It is observed that one or two extrema at high energies
($E \approx$ $40$--$60$ MeV) may still survive for the Livermore model
where the spectrum extends to higher energies, but their 
clean identification
would require $\sim 10^5$ events at a scintillation
detector in a single time bin.
Figure~\ref{fig-flux}(b) also shows the positron energy spectrum
for $10^5$ events, binned in $0.5$ MeV energy intervals, which is
approximately the energy resolution of a scintillation
detector near $E = 40$ MeV.

The total number of events expected even at a future
50 kt scintillation
detector is $\sim 10^4$ for a SN at a distance of $10$ kpc \cite{lena}.
Even with 1 sec time bins, the number of events in each bin
will be $\sim 10^3$. This is a number too small for our
purposes. Thus, the identification of the phase effects seems
very unlikely, unless the SN is as close as a kpc.

The density profiles in the shock wave are quite uncertain,
and one may expect that some possible profiles give rise to
oscillations with larger amplitudes and wavelengths, which 
would be easier to observe.
The amplitude of oscillations is proportional to the ratio
$|A'(T)/A'(R)|$ [See (\ref{a-def}) and (\ref{master-eqn})].
Therefore, larger amplitudes need a larger ratio $|A'(T)/A'(R)|$ 
whereas larger 
wavelengths need a sharp tail, i.e. a larger $|A'(T)|$. 
At the same time, the adiabaticity parameter at the two 
resonances has to be $0.1 \lsim \gamma \lsim 2$ for
$E \approx 40$ MeV.
However, we find that even with such finely tuned density profiles,
the improvement in observability is not significant.

\section{Concluding remarks}
\label{concl}

When the neutrinos escaping from the core of a core collapse
SN pass through the shock wave, they may encounter multiple
``$H$'' resonances corresponding to $\dmsq_{\rm atm}$ and $\theta_{13}$,
when the shock wave is in the regions with densities  around 
$500$--$5000$ g/cc.
We have shown that this necessarily gives rise to oscillations in the 
neutrino survival probabilities, which we have calculated 
as a function of energy.
We present an analytical approximation for small mixing angles
and show that 
the oscillations are a significant effect: they can be of the same 
order as the non-oscillating terms.
The typical values of $\theta_{13}$ that gives rise to the 
oscillation features are in the ``transition region''  of the
neutrino mixing parameter space
\cite{ds} that is significant in range
($10^{-5} \lsim \sin^2 \theta_{13} \lsim 10^{-3}$) 
but is usually neglected in the SN analysis for the sake of
simplicity.

The local maxima and minima in the survival probabilities
of $\nue$ or $\nuebar$ are determined by
the relative phase accumulated by the neutrino mass eigenstates
between the multiple resonances.
The positions of these extrema depend on the density profile and 
are independent of the primary neutrino spectra.
If these extrema were identified they would
reveal information on the propagation of the shock wave:
its location as well as the density variation present around it.
Since neutrinos are the only particles that can carry
information about the shock wave while it is still 
deep inside the exploding star, it is important to explore 
the observability of these phase effects. Moreover, the mere identification of these effects in the $\nuebar$ spectrum would establish the inverted hierarchy and
nonzero $\theta_{13}$, which are two of the most important
quantities in neutrino phenomenology, and even faint chances of their
determination should be explored.
 
It is interesting that for typical shock wave profiles,
oscillations with ``wavelengths'' of
$2$--$8$ MeV are indeed present in the neutrino survival probabilities
in the energy range $E \approx$ $30$--$60$ MeV.
These wavelengths are tantalizingly close to the
resolving power of the neutrino detectors.
However, the inability of the 
detectors in reconstructing the energy of the incoming neutrino 
tends to wash out the oscillation pattern. 
For typical shock wave density profiles,
the energy resolution of water Cherenkov detectors is insufficient 
to detect the oscillations.
Even a scintillation detector with a superior energy 
resolution will need $\sim 10^5$ events in the relevant time bin
of $4$--$5$ sec postbounce for identifying one or two extrema in the most optimistic
scenario. 

 Therefore, we expect that neglecting the phase effects is a valid practical approximation except under extreme cases. Note that it is still possible to observe other robust signatures of shock wave propagation like the dips in the time spectrum of number of events \cite {fogli} or the double dip feature in $\la E \ra$~\cite{revshock}. Some of these features survive even in the presence of somewhat extreme stochastic density fluctuations, as discussed in \cite{stochastic}. The oscillatory effects might contribute additional scatter to these signals but will not spoil the signatures.  

We must qualify the above conclusions by stating the effects ignored in our analysis to obtain the present results.  We have assumed a smooth spherically symmetric density profile and ignored anisotropies that are likely to be present \cite{revshock}. This is a justified assumption only as long as the deviations from this assumed profile occur only on transverse distances larger than the size of the neutrinosphere so that the neutrinos in our line of sight do not experience the anisotropy. We have not included the recently discussed collective effects of coherent flavor development \cite{coll-duan1, coll-raff}, which may be important in the inverted hierarchy. However, the extent and nature of its impact on flavor conversion has not been worked out in detail, which makes it difficult to be included in the present analysis. Similarly, the effects of a realistic spectrum of stochastic fluctuations in the medium density or turbulent convections behind the shock wave are yet to be calculated for SN neutrinos \cite{stochastic, noise}. We think that with a better understanding of the collective effects and density fluctuations one could include their effect on the realistic observability of the phase effects.

The phase effects pointed out here result from the
interference between two or more MSW resonances. This phenomenon is not restricted to the SN alone, but may occur whenever neutrinos pass through nonmonotonic matter densities
and the resonances are semiadiabatic.
The technique developed in this paper for treating coherent
multiple resonances is applicable to such cases.

\section*{Acknowledgments} 

We would like to thank R. Buras and L. Scheck for help in 
understanding features of the shock wave.
AD would like to thank 
the Max Planck Institute for Physics
for hospitality during the initial part of the work.
We would like to especially thank B. Bhattacharya for useful
discussions.
We are grateful to S. Choubey, H. -Th. Janka, E. Lisi, 
G. G. Raffelt and R. Tom\`as for comments on the manuscript.
This work was partly supported through the Partner Group
program between the Max Planck Institute for Physics and
Tata Institute of Fundamental Research.


\end{document}